\documentclass[conference]{IEEEtran}
\IEEEoverridecommandlockouts
\ifCLASSINFOpdf
   \usepackage[pdftex]{graphicx}
  \graphicspath{{./figures/}}
\else
\fi
\usepackage{amsmath}

\usepackage{soul}

\clearpage

\DeclareMathOperator*{\argmax}{arg\,max}
\begin{document}
%

\title{A Multi-Agent Neural Network for Dynamic Frequency Reuse in LTE Networks}

\author{\IEEEauthorblockN{Andrei Marinescu, Irene Macaluso, Luiz A. DaSilva}
\IEEEauthorblockA{CONNECT, 
Trinity College Dublin\\
Email: \{\textit{marinesa, macalusi, dasilval}\}@tcd.ie
}
}


\maketitle

\begin{abstract}
Fractional Frequency Reuse techniques can be employed to address interference in mobile networks, improving throughput for edge users.
There is a tradeoff between the coverage and overall throughput achievable, as interference avoidance techniques lead to a loss in a cell's overall throughput, with spectrum efficiency decreasing with the fencing off of orthogonal resources.
In this paper we propose MANN, a dynamic multi-agent frequency reuse scheme, where individual agents in charge of cells control their configurations based on input from neural networks. The agents' decisions are partially influenced by a coordinator agent, which attempts to maximise a global metric of the network (e.g., cell-edge performance). Each agent uses a neural network to estimate the best action (i.e., cell configuration) for its current environment setup, and attempts to maximise in turn a local metric, subject to the constraint imposed by the coordinator agent.
Results show that our solution provides improved performance for edge users, increasing the throughput of the bottom 5\% of users by 22\%, while retaining 95\% of a network's  overall throughput from the full frequency reuse case. Furthermore, we show how our method improves on static fractional frequency reuse schemes.

\end{abstract}

\section{Introduction}

In orthogonal frequency-division multiple access (OFDMA) mobile networks, where spectrum efficiency increases with frequency reuse, cell-edge performance is affected by increased interference between neighboring cells in edge areas. To address this, Fractional Frequency Reuse (FFR) schemes have been proposed to improve cell-edge performance. These allocate orthogonal blocks of the available spectrum between neighbouring cells, in order to decrease interference. One such scheme is partial frequency reuse (PFR), which is also known as fractional frequency reuse with full isolation \cite{hamza2013survey}, or as strict frequency reuse. PFR divides the bandwidth into a common block that is to be used by center users, with full reuse between cells (i.e., reuse factor of 1), and another group of 3 orthogonal blocks (i.e., with reuse factor of 3), so that edge users in neighbouring cells operate in orthogonal bands.  As such, the reuse factor is greater than 1 (the factor of full frequency reuse), and lower than 3 (the factor for hard frequency reuse, where the spectrum is evenly divided into three orthogonal chunks for three neighbouring cells/sectors).

In PFR, there are three types of parameters that can be controlled:

\begin{itemize}
\item proportion of bandwidth allocated to center users vs. edge users;
\item threshold defining center vs. edge users (e.g., SINR level or RSRQ - reference signal received quality);
\item power allocated to the center users' block and/or edge users' block.
\end{itemize}

However, schemes such as PFR typically come at the cost of overall cell capacity \cite{andrews2011tract}, and in some cases, a static FFR scheme can even be detrimental to cell-edge users. Dynamic FFR schemes that can adapt to environment conditions (e.g., traffic load and UE distributions) are more suitable in such instances \cite{hamza2013survey,gonzalez2010performance}. 

With dynamic schemes, inter-cell coordination allows for more efficient utilization of frequency resources between cells, at the cost of communication overhead and increased computational complexity. In dynamic environments, the problem of optimally allocating resources in order to maximise a metric becomes NP-hard \cite{shams2014survey,necker2011novel}, therefore infeasible to solve through centralised algorithms when many variables are involved. For this, heuristic algorithms need to be employed to obtain a solution in a timely manner.

In this paper we propose MANN, a  Multi-Agent Neural Network solution which allows cells to dynamically select the best fractional frequency reuse parameters for their current environment conditions, namely the geographical distribution of active subscribers and the amount of traffic generated. 
 



Other attempts have been made to dynamically address cell interference through FFR techniques and thus improve the performance experienced by cell-edge users. In \cite{rahman2010enhancing}, a dynamic scheme based on inter-cell coordination is proposed, which relies on a central controller to compute the optimal frequency allocations by solving a binary integer optimization, which is an NP-complete problem \cite{karp1972}. This solution is only tested considering uniform distributions of UEs, on average 11 per cell, and with high communication overhead and additional channel interferer information required from each UE. A heuristic solution was later proposed in \cite{yu2013multi}, to address the issue of computational complexity. The solution is an interatively updating algorithm that at best converges to local optima. It can be implemented both centrally and in distributed fashion, but still involves high communication overhead, which might deem it impractical for actual network deployments. In \cite{bernardo2010rl}, a centralised solution with a reinforcement learning (RL) based network level controller is proposed to allocate orthogonal channels to cells so as to satisfy a minimum level of throughput per UE. However, this solution is susceptible to poor performance in situations when it is forced to explore, as the RL algorithm is trained only on uniform geographic distributions of UEs per cell, and will need to learn how to address various types of clustered UE patterns it may encounter.

Self-organising techniques have also been employed to dynamically address a similar problem, in particular soft FFR. In soft FFR, the full spectrum is used by all cells, but on a third of the spectrum edge users are served with higher power levels. Neighbouring cells serve their edge users on orthogonal parts of the spectrum.
For this scheme, an iterative multi-sector gradient algorithm is proposed in \cite{stolyar2009selforg} to address interference between cells, which gradually actuates different sub-band power levels to maximise network 
utility. While more spectrum efficient, soft FFR does not provide the same benefits for edge users as FFR schemes with exclusive orthogonal blocks. 

\par
The solution we propose in this work dynamically addresses the problem of fractional frequency reuse through a semi-distributed architecture, with minimal exchange of information between a cell agent and the coordinator agent. The cell agents are in charge of dynamically controlling FFR parameters in their sectors and of reporting estimated gains under each possible bandwidth to the coordinator agent, based on regression through neural networks. The coordinator agent aggregates local performance estimates from the underlying cell agents, and after evaluating these it broadcasts back to cell agents the best global orthogonal division of spectrum between edge and center users, i.e., the bandwidth configuration that best avoids interference between cells while maximising global network performance. Once a global orthogonal bandwidth division is imposed by the coordinator agent, cell agents are in charge of optimising their local PFR configuration to best suit their own users' geometry.


We tested our solution on a broad set of UE distributions, which involve a large variation in user patterns: from uniform to clustered; with sparse or dense clusters; with cell edge or cell center clusters; and with a wide ranging number of UEs per cell (from 0 to 70 UEs per cell). The next section describes our proposed solution.

%

\section{Proposed Solution}
\label{sec:methodology}

MANN, our proposed solution, is a dynamic PFR scheme with coordination. 
Here, we focus on controlling the \textit{bandwidth} allocated to center users vs. edge users (i.e., which proportion of the spectrum is to be shared by all cells and which is orthogonally divided among neighboring cells) and the \textit{RSRQ threshold} defining edge users. The power level is kept fixed, with a higher power level allocated for edge users than for center users.

\subsection{Multi-Agent System Architecture}

\begin{figure}[!htb]
\centering
\includegraphics[width=\linewidth]{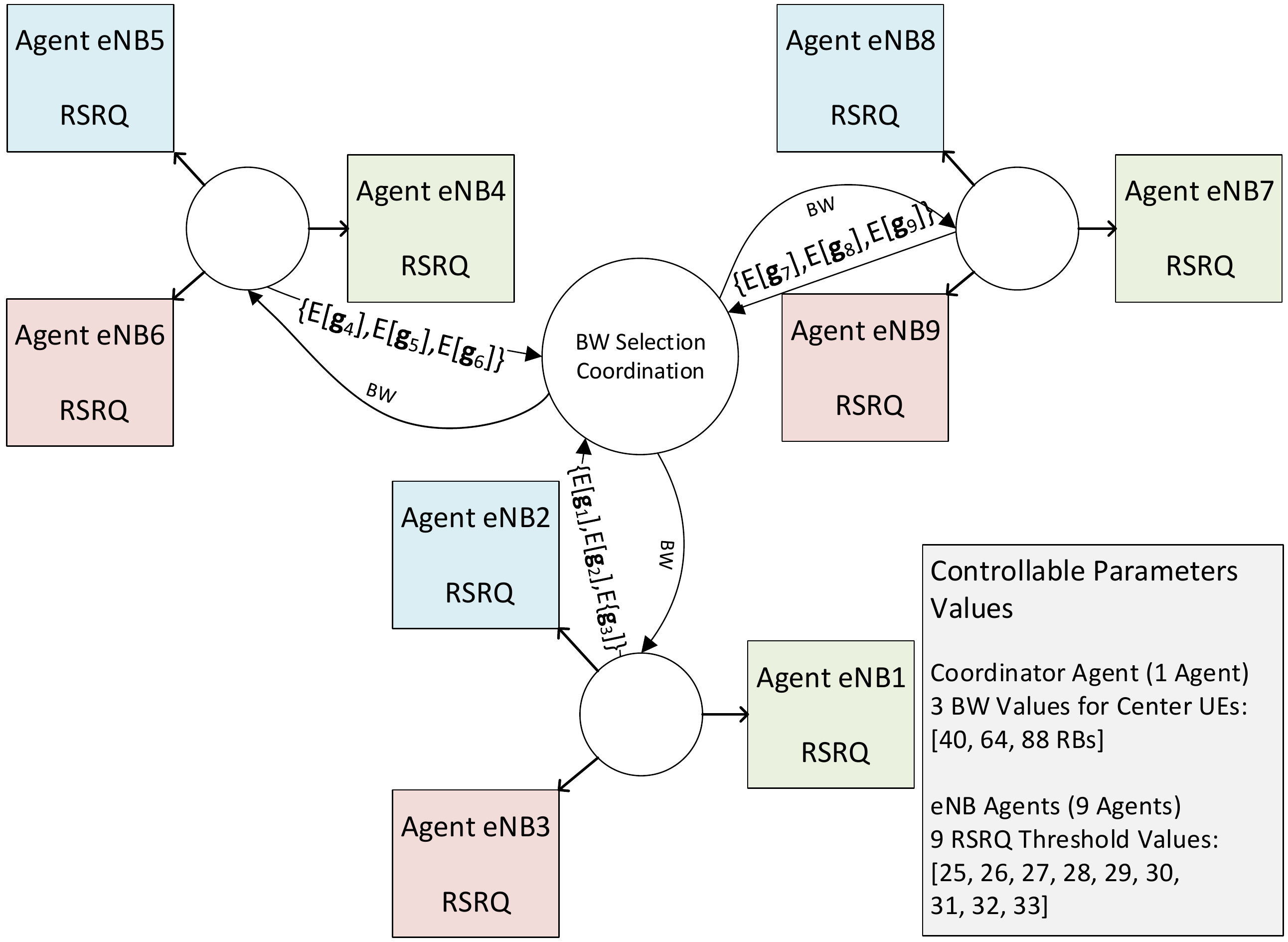} 
\caption[c]{Multi-agent system architecture. 
}
\label{fig:mas_architecture}
\end{figure}

The system architecture is shown in Fig. \ref{fig:mas_architecture}. In MANN, agents are placed in control of each cell, and are only partially influenced by a coordinator agent, thus operating in a semi-distributed manner. The coordinator agent is responsible for resolving potential interference generating conflicts between neighbouring cells, and for deciding how to divide the bandwidth between edge users and center users based on maximising a global metric of performance. Each agent sends a forecast estimate of maximum local performance attainable for each bandwidth selection possible, noted as $E[g_{i}]$ in Fig. \ref{fig:mas_architecture}, and this information is aggregated at the coordinator agent's side. The bandwidth setup with the highest global performance is then selected by the coordinator agent and broadcasted to the cell agents ($BW$ in Fig. \ref{fig:mas_architecture}). These in turn will use the imposed bandwidth to select the most suitable edge/center user threshold with respect to their own user geometry. Note that the coordinator agent can either reside in one of cells or can be a higher level entity in the network. 

\subsection{Cell Agent's Actions}


Each cell agent can control both RSRQ thresholds (where any UE with an RSRQ above this threshold will be considered as a center UE instead of an edge UE) and bandwidth selection. When coordination is involved, all agents report to the coordinator agent the estimated best performance attainable under each bandwidth selection. The coordinator agent aggregates these estimates and broadcasts back the bandwidth that maximises the global agents' performance under the metric of choice selected for the network. 

The cell agents adopt the bandwidth selected by the coordinator, to avoid potential edge user interference with neighbouring cells. Afterwards, they individually choose the RSRQ thresholds that provide the highest gains given their current UE distributions and the  bandwidth restriction imposed by the coordinator agent. The cell's decision is aided by a neural network that resides at cell agent level. 

\subsection{Neural Network Environment Abstraction and Performance Forecasting}

An estimate of the local performance metric is obtained by each cell agent based on the regression performed by a neural network. We chose a neural network for this purpose, since the performance metric is based on throughput achievable by UEs, and thus is a non-linear function that depends on pathloss, noise, interference, user geometry, and bandwidth available, among others. The neural network uses as input, in normalized form (values scaled to [0,1] interval):

\begin{itemize}
\item 10 input neurons with information about the environment, 1 per RSRQ bin (i.e., number of UEs with RSRQ in each interval/bin [$<$25], [25-26], ..., [32-33], [$>$33]);
\item 1 input neuron for the bandwidth action (i.e, 40, 64 or 88 RBs exclusively used for center UEs);
\item 1 input for the RSRQ threshold (i.e., 25,26, ..., 33).
\end{itemize}

\begin{figure}[!htb]
\centering
\includegraphics[width=0.5\linewidth]{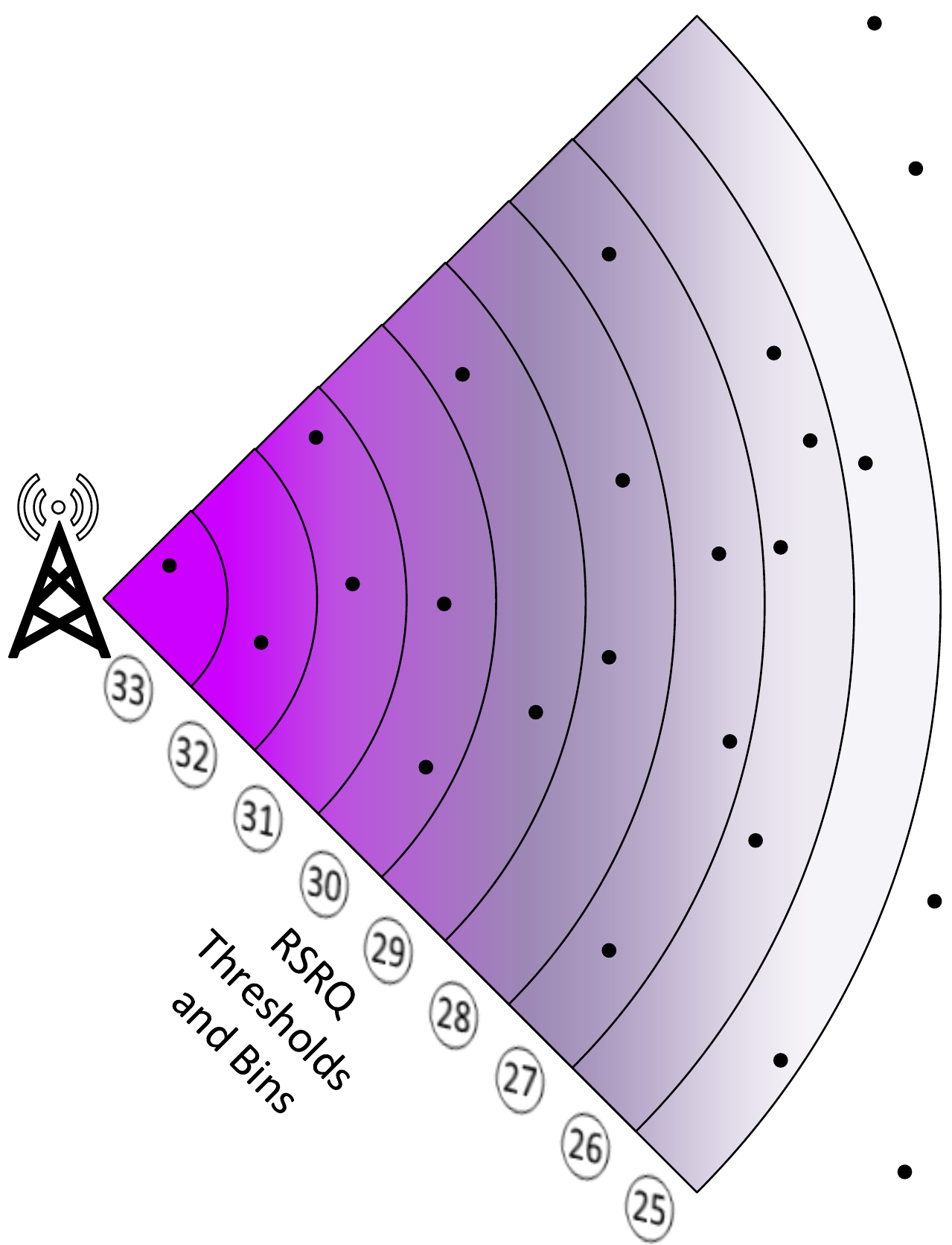} 
\caption[c]{Cell agent and UE geometry abstraction.}
\label{fig:rsrq_cell}
\end{figure}

We abstracted features of the UE geometry that characterize the cell environment in the form of number of UEs per RSRQ bin, as pictured in Fig. \ref{fig:rsrq_cell}, where for example in bin [25-26] there are 2 UEs. Afterwards, we used 5 hidden layers\footnote{We empirically reached this architecture, as from 5 layers upwards the neural network is sufficiently complex to accurately estimate most performance metrics/functions based on UE throughput. Any further increase in number of layers did not bring additional performance improvements.}, each of 32 fully connected neurons, to be able to extract and process the information from the environment. Finally, these link to an output neuron which provides an estimate of the performance metric, considering the current environment state and a combined bandwidth-RSRQ threshold action choice. 

In total 27 actions are possible (3 for BW x 9 for RSRQ), therefore 27 possible performance values for each state\footnote{Note that the state-space can still be quite large despite the abstraction; for example if we consider up to 4 UEs per bin, for a maximum of 40 UEs per cell, we have approximately $10^{7}$ possible states, and the cardinality of the action-state space thus becomes $27 \times 10^{7}$.}. The neural network is trained through supervised learning based on a set of observed inputs and outputs, where the output is the performance metric of choice. The performance metric is a function of the throughput observed for all UEs in a cell, e.g., mean throughput, harmonic mean throughput, minimum throughput, maximum throughput, etc. The desirable performance metric needs to be selected before beginning the training process, and then the outputs are automatically configured in accordance to the metric for the actual training stage. Once the network is trained, it is used by cell agents to take decisions. The training process needs to be done on a large and diverse enough set of states and action combinations to avoid high bias and overfitting.

The next section describes the scenario we have chosen for evaluating the proposed algorithm.

%
%
%

\section{Evaluation Scenario}
\label{sec:scenario}

\begin{figure}[!htb]
\centering
\includegraphics[width=\linewidth]{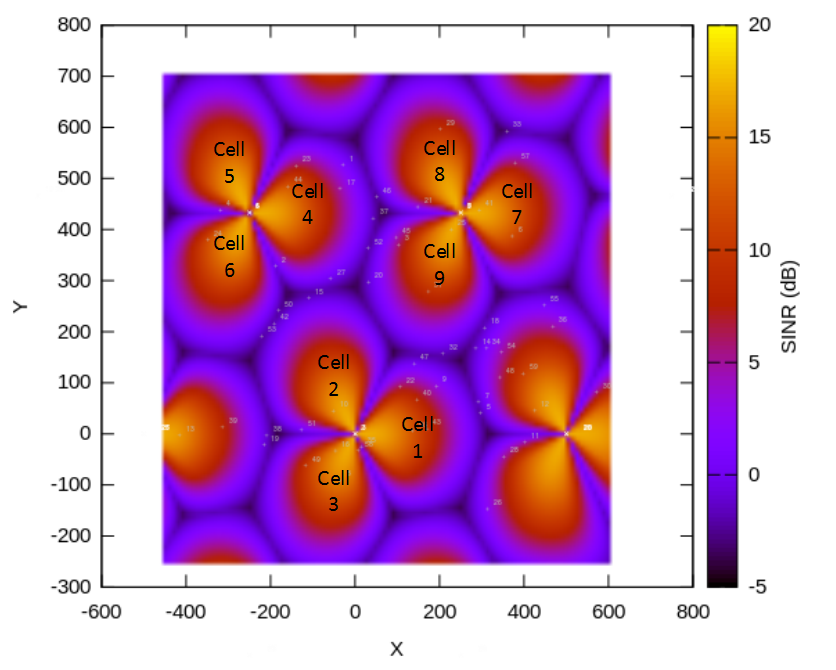} 
\caption[c]{Cell setup in ns-3: inner 9 cells.}
\label{fig:9cells}
\end{figure}

We have evaluated our algorithm using ns-3's LTE simulator \cite{baldo2011open}. The configuration of the cells is done in accordance to 3GPP TR 36.814 \cite{3gpp36814}, using a validated setup for which further details can be found in our previous work \cite{marinescu2017ns3}. Our setup uses 36 cells, with 27 used as a wrap-around ring around the inner 9 cells. The proposed solution is implemented on top of the 9 inner cells (3 sites, each with 3 cells), as can be seen in Fig. \ref{fig:9cells}. 

For this, we have modified ns-3's existing Strict Frequency Reuse scheme from the LTE module, which represents an implementation of partial frequency reuse with full isolation. While the ns-3 scheme is static, with a default configuration involving fixed bandwidths and RSRQ thresholds (28 RBs for common bandwidth, and RSRQ threshold of 25), our modified version operates in a dynamic manner, being able to reconfigure the network at every time-step (in this case every 1 second of real-time), with additional configurations implemented for bandwidth and RSRQ threshold to support this.

To reduce the complexity of the decision and the communication overhead, we consider only three options of bandwidth configurations, in an LTE network with a 20 Mhz band, where 100 RBs are used for downlink (e.g., for case with 40 RBs allocated to center users, a cell will have an exclusive orthogonal block of 20 RBs of the remaining bandwidth to use for its own edge users). For RSRQ threshold control we consider 9 options, integers from the [25-33] interval, in accordance to Section 9.17 of 3GPP TS 36.133 \cite{3gpp36133}. 

\subsection{User Distributions}

For the training process we have generated a set of UE distributions based on stochastic geometry models.
We employed R and used the Thomas cluster process (a special case of Poisson Point Process) to create UE distributions, which were then imported in ns-3. 
Such a process better models the clustering of users expected in real mobile networks.
First, we have generated a set of 40 UEs randomly uniformly distributed across the frame from Fig. \ref{fig:9cells}. Afterwards we used the following parameters to drop additional UEs on the initial uniform layer:

\begin{itemize}
\item Number of macro clusters centered on the 3 sites: \{3\};
\item Number of micro clusters per macro cluster: \{3\};
\item Radius of micro cluster: \{75, 100, 125\} meters;
\item Number of users per micro cluster: \{11, 12, ..., 20\};
\item Displacement of micro cluster from macro cluster site location:  \{50, 100, 150, 175, 200, 250\} meters.
\end{itemize}


In total there are 180 possible configurations (i.e., $1 \times 1 \times 3 \times 10 \times 6$), with the numbers of UEs per cell in the [0-70] range. However, for each configuration every generated instance will be different from each other, due to the randomness involved when generating the location of clusters and of users within a cluster. 

\subsection{Training and Testing the Neural Network}

For the training stage of our algorithm we have generated 4 instances for each of the 180 configurations, for a total of 720 UE distribution scenarios. We let the cells perform 30 random actions over 30 seconds of real-time in each distribution (one per second of real-time), and then we store the obtained results for each action. Since we have 9 cells and 30 actions taken per cell, we have obtained approximately $2\times10^{5}$ input/output samples for training the neural network.

For the testing stage, we have generated an additional instance for each of 180 configurations, and explored again 30 random actions per cell, thus obtaining an additional 50000 samples to evaluate the accuracy of the trained neural network\footnote{Acquiring the training and test samples was the most computationally expensive process, as it took approximately two weeks of running parallel ns-3 simulations on a 24 core server to collect the necessary data.}.

The neural network is trained using Keras on top of Tensorflow \cite{chollet2015keras}. Training is done on the shuffled $2\times10^{5}$ samples, using a mini-batch size of 50 samples and for 30 epochs, where 20\% of the samples are employed for cross-validation\footnote{The training process takes approximately 2 minutes on a Core i7 CPU with four 3.8 Ghz cores.}. The neural network layers use tanh and sigmoid activation functions, and the gradient-based Adamax optimizer \cite{kingma2014adamax}.

\subsection{Local and Global Performance Metrics}
\label{sec:perf_metrics}

We have chosen a metric with the intention of improving the performance for edge UEs, in order to evaluate our proposed solution. At \textbf{local} cell level, the performance metric is represented by the minimum throughput achieved by any UE in the cell, weighted by the total number of UEs:

\begin{equation}
\label{eq:mincell}
MetricCell_{j} = n_{j} \times \min(TPue_{j1}, TPue_{j2},.., TPue_{jn})
\end{equation}

where $TPue_{ji}$ represents the throughput achieved by UE $i$ of cell $j$, and $n_{j}$ is the total number of UEs in cell $j$. Once the neural network is trained to perform regression on this metric, the agent chooses the RSRQ threshold that maximizes Eq. \ref{eq:mincell} for each possible value of bandwidth based on the current environment condition:

\begin{equation}
\label{eq:cellmaxmin}
DesiredAction_{j}= \underset{A_{BW_{k}RSRQ_{l}}}\argmax MetricCell_{j}
\end{equation}

where $A_{BW_{k}RSRQ_{l}}$ represents a combined action composed of bandwidth action $k$ from the $\{40,64,88\}$ set and RSRQ threshold action $l$ from the $\{25,26, ..,33\}$ set.

Each cell agent will provide an estimate of the maximum achievable gain considering the current metric for each of the three available bandwidth choices, under the form of a vector:
\begin{equation}
E[g_{cell_{j}}] = <\underset{BW_{40}}E[g_{cell_{j}}], \underset{BW_{64}}E[g_{cell_{j}}], \underset{BW_{88}}E[g_{cell_{j}}] >
\end{equation}

where $\underset{BW_{k}}E[g_{cell_{j}}]$ represents the maximum attainable gain for cell $j$ under bandwidth choice $k$.

Once the coordinator agent receives the estimated gain vectors from all the cells under its influence, it will compute the highest total gain achievable for each bandwidth, which represents the \textbf{global} metric at coordinator agent level: 

\begin{equation}
\label{eq:coordmax}
WinningBandwidth = \underset{BW_{k}}\argmax \underset{j={1..m}}\sum\underset{BW_{k}}E[g_{cell_{j}}]
\end{equation}

where $m$ is the total number of cell agents coordinated, in this scenario equal to 9. The coordinator agent then imposes the winning bandwidth by broadcasting it to all the cell agents under its influence, as can be seen in Fig. \ref{fig:mas_architecture}. This bandwidth will be then fixed for each cell to avoid interference, while each cell agent will afterwards select the RSRQ threshold that maximises its own gain under the selected winning bandwidth.

The next section presents the results we have obtained in the scenario described above, considering the maxmin performance metric, i.e., maximizing the minimum throughput experienced by any UE in the cell.

\section{Results and Analysis}
\label{sec:results}

\subsection{Neural Network Performance}

We first evaluate the accuracy of the neural network's forecasts of the desired performance metric, in this case the weighted minimum throughput of a cell, as described in Section \ref{sec:perf_metrics}. After training the neural network, we have verified its performance on the separate testing set comprising 180 UE drops. The results obtained are shown in Table \ref{tab:nn_results}, and are based on normalised values within the [0,1] interval. The table includes the following metrics: correlation, mean absolute error (MAE), mean absolute percentile error (MAPE), and root mean square error (RMSE).

\begin{table}[!htb]
\centering
\caption{Performance of Neural Network Prediction for Minimum Throughput Prediction}
\label{tab:nn_results}
\begin{tabular}{|l|l|l|l|l|}
\hline
\textbf{Metric Type} & \textbf{Correlation} & \textbf{MAE} & \textbf{MAPE}  & \textbf{RMSE}\\ \hline
\textbf{Forecast}                & 0.92                 & 0.020         & 0.029         & 0.034\\ \hline
\end{tabular}
\end{table}

As can be noticed from Table \ref{tab:nn_results}, the neural network is able to accurately forecast the minimum throughput that will be achieved by a cell for every action under the given UE distributions, obtaining a forecasting error of 0.029 MAPE. 

\subsection{Multi-Agent Frequency Reuse Performance}

We employ the trained neural network in our multi-agent setup, and use it to assist the dynamic adjustment of frequency reuse parameters for each of the 9 inner cells, according to the process described in Section \ref{sec:methodology}. 

We compare the performance obtained with our solution against a baseline where full frequency reuse is employed with a proportional fair scheduler, i.e., where all the 100 RBs are used in each cell for all UEs. 

The experiments were performed over the 180 UE distribution instances from the test set. While coordination and actions are performed by all 9 cells, only results from the 3 inner cells (Cell 2, 4 and 9 from Fig. \ref{fig:9cells}) were considered in the evaluation, as unlike the outer cells these are under the influence of dynamic coordination effects of all of their neighbours. 

\begin{figure}[!htb]
\centering
\includegraphics[width=\linewidth]{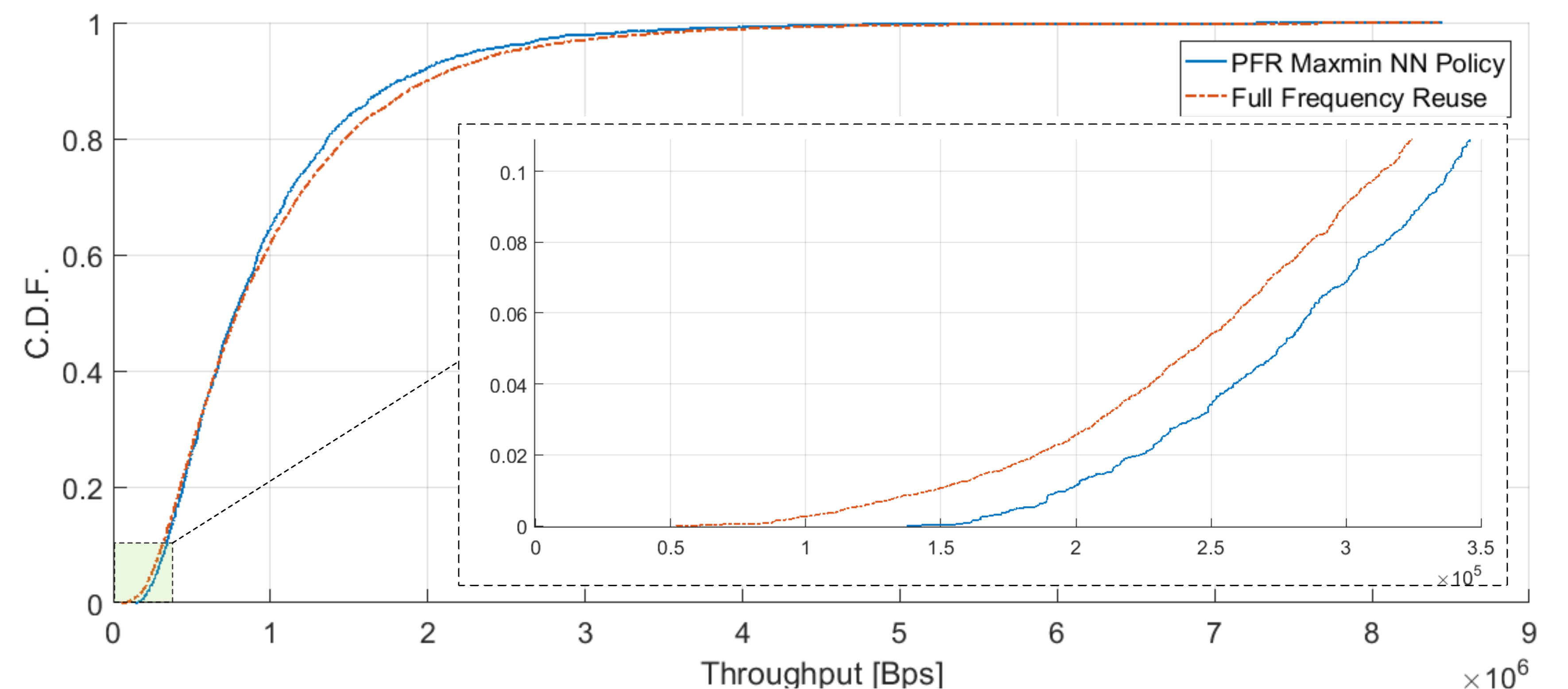} 
\caption[c]{MANN vs. Full Frequency Reuse: overall and bottom 10\% UE throughput.}
\label{fig:ml_vs_fr}
\end{figure}

Fig. \ref{fig:ml_vs_fr} shows the UEs throughput cumulative distribution functions (CDFs) for the baseline, full reuse, and our proposed solution. 
In it we also provide a zoom-in over the bottom 10\% UEs, as for this evaluation the purpose is to increase the performance of the cell-edge UEs. While the chosen performance metric specifically addresses the maximisation of the minimum achievable throughput by each user, our expectation was that it would also reflect in improved performance for cell-edge UEs. Further results are presented in Table \ref{tab:mas_performance}.

\begin{table}[!htb]
\centering
\caption{MAS Frequency Reuse with MANN vs. Baseline (Full Reuse with Proportional Fair Scheduler) }
\label{tab:mas_performance}
\begin{tabular}{|l|l|l|l|l|}
\hline
\textbf{\begin{tabular}[c]{@{}l@{}}Performance \\ Metric\end{tabular}}          & \textbf{\begin{tabular}[c]{@{}l@{}}Bottom \\ 10\% UEs\end{tabular}} & \textbf{\begin{tabular}[c]{@{}l@{}}Bottom \\ 5\% UEs\end{tabular}} & \textbf{\begin{tabular}[c]{@{}l@{}}Bottom \\ 1\% UEs\end{tabular}} & \textbf{\begin{tabular}[c]{@{}l@{}}Worst \\ UE\end{tabular}} \\ \hline
\textbf{\begin{tabular}[c]{@{}l@{}}Improvement over \\ Full Reuse\end{tabular}} & 15\%                                                                & 22\%                                                               & 56\%                                                               & 264\%                                                        \\ \hline
\end{tabular}
\end{table}

\begin{figure}[!htb]
\centering
\includegraphics[width=\linewidth]{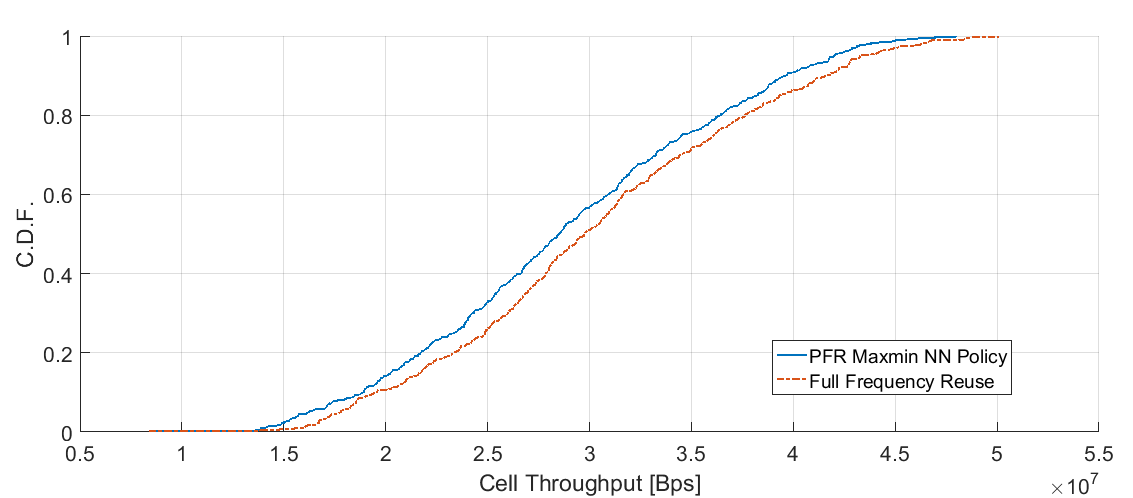} 
\caption[c]{MANN vs. Full Frequency Reuse: cell throughput.}
\label{fig:ml_vs_fr_cell}
\end{figure}

The most significant gain occurs in terms of the minimum throughput achieved by all users, where a 264\% improvement can be noticed. Moreover, the improvement for cell-edge UEs (here considered to be the bottom 5\%, in accordance to 3GPP TR 36.814 \cite{3gpp36814}) is still significant, providing a gain of 22\%. In fact, our solution outperforms the baseline for the bottom 33\% UEs while also retaining over 95\% of the baseline's total throughput achieved, both at cell and UE level. The cells' throughput performance is shown in Fig. \ref{fig:ml_vs_fr_cell}.

\begin{figure}[!htb]
\centering
\includegraphics[height=7cm, width=\linewidth]{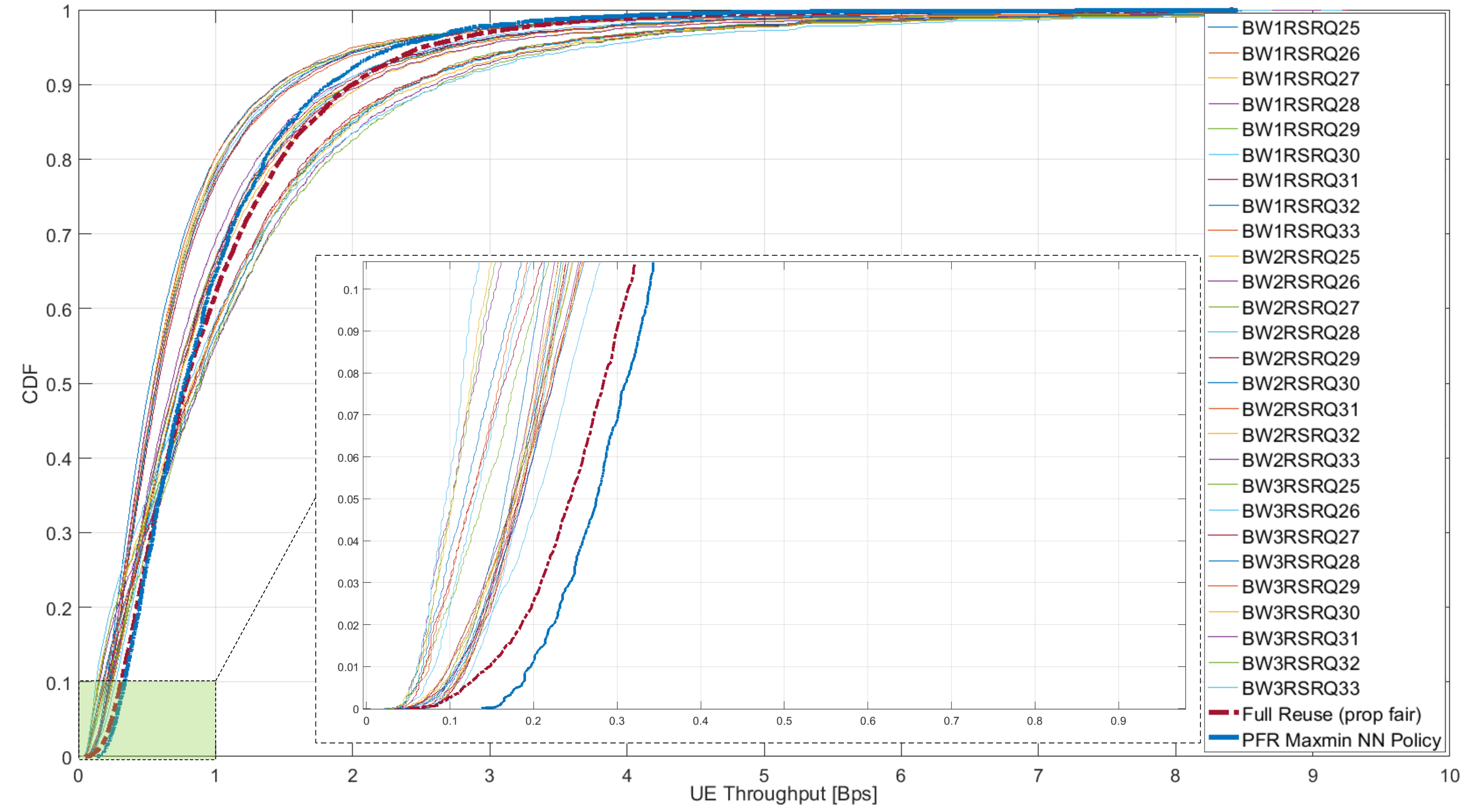} 
\caption[c]{MANN vs. baseline and all possible PFR configurations.} 
\label{fig:ml_vs_all}
\end{figure}

In addition, Fig. \ref{fig:ml_vs_all} presents the same results shown in Fig. \ref{fig:ml_vs_fr} alongside the 27 possible static configurations. Interestingly, on average from the 180 UE drops, it can be noticed that no static configuration outperforms the baseline (full reuse) for the cell-edge users. While there are certain UE drops where for a particular cell a static PFR configuration provides gains compared to the baseline, these gains are not reflected when considering a network-wide scenario with multiple cells that use the same configuration, as the configuration can actually be detrimental to its neighbours, and even more so when the UE geometry changes. Moreover, in our scenario, we noticed that even hard frequency reuse (the most edge UE friendly FFR scheme), shows gains for at most the 0.8\% bottom UEs with respect to throughput, while retaining only 61\% of the baseline's total throughput.

\section{Conclusions and Future Work}
\label{sec:concl}

We have presented an autonomous multi-agent neural network based control scheme that enables efficient dynamic frequency reuse control in LTE networks with low communication overhead. We showed how our solution outperforms both static frequency reuse configurations and the full frequency reuse baseline in terms of cell-edge UE performance, without incurring significant overall network throughput losses. We have noticed, in fact, that in our scenarios static frequency reuse schemes do not achieve better performance for cell-edge users when compared to full frequency reuse with a proportional fair scheduler. These results were obtained on an average of 180 different UE distributions across 9 cells, comprising various types of clustered UE patterns (e.g., sparse, dense, close to cell sites, far from cell sites, with low number of UEs, high number of UEs, etc.).

Moreover, the proposed multi-agent architecture and neural network solution can be employed to also maximise other types of metrics at cell level, such as mean UE throughput. We have additional results for this metric, where our solution achieves an increase of 32\% in overall network throughput compared to full frequency reuse, however at the cost of decreased cell-edge performance\footnote{These results are not included in the paper due to space constraints.}. No change in the multi-agent architecture or neural network was required for this, as the neural network can be trained on almost any type of metric and will still be able to accurately forecast the performance\footnote{The neural network has been successfully tested on other throughput-related metrics such as geometric mean, logarithmic mean, harmonic mean, etc.}. Since the structure of the neural network remains the same, weights for the network trained under other types of metrics can be directly loaded into the multi-agent system to automatically adjust its behaviour in accordance to the desired metric to be maximised.

The proposed multi-agent coordination mechanism could also be considered as a hierarchical system, where there can be coalitions of agents, each coordinated by a separate coordinator agent, which in turn reports to a higher-level coordinator entity to maximise a global network metric. In this work the network level metric at coordinator agent level was set to maximise the aggregated sum of minimum throughput achievable per cell. Other metrics can be employed at this level as well, such as maximising the social welfare of agents, as there can be cases where maximising the global metric can be very detrimental to some of the underlying cell agents.

For the next steps, we consider allowing the neural network to learn online by performing gradient descent steps on each newly collected sample, so as to better adapt to potential unencountered situations where it might underperform. In addition, we are interested in observing the performance of the system in bursty traffic conditions, where instead of numbers of UEs per RSRQ bin the number of bytes to be sent per bin becomes more relevant. Another parameter that could be controlled in addition to bandwidth and RSRQ actions would be the power levels allocated to edge and center UEs. This opens up other possibilities for coordination between agents to mitigate potential interference.


\section*{Acknowledgment}

This work was partly funded by Huawei Sweden and the Science Foundation Ireland, the latter under Grant number 13/RC/2077. 
We gratefully acknowledge discussions and assistance from Gunnar Peters, at Huawei, and Jacek Kibilda, at CONNECT, in the course of this work.

\bibliographystyle{IEEEtran}
\bibliography{biblio}

\end{document}